\documentclass[aps,prd,superscriptaddress,twocolumn,
	amsfont,
	amssymb,
	amsmath,
	multirow,
	notitlepage,
	longbibliography,
	nofootinbib]{revtex4-1}
\usepackage[colorlinks=true, 
	linkcolor=blue, 
	citecolor=blue, 
	urlcolor=blue, 
	linktocpage=true]{hyperref}
\usepackage[english]{babel}
\usepackage{graphicx}

\begin{document}

\title{Universal scaling law for quantum droplet formation}
\date{\today} 

\author{Ian G. Moss}
\email{ian.moss@newcastle.ac.uk}
\affiliation{School of Mathematics, Statistics and Physics, Newcastle University, 
Newcastle Upon Tyne, NE1 7RU, UK}

\begin{abstract}
Given the right set  of circumstances,  ultracold quantum gases are able to change character
and condense into a liquid state of quantum droplets.  The size distribution of the droplets is 
determined dynamically in the condensation  process.  A semi-quantitative argument is presented 
which suggests that,  at zero temperature,  a multiple droplet system has is a preferred scale 
$\propto v^{-1/3}$,  where $v$ is the rate of change of parameters at the 
time of droplet formation.  Numerical simulations of two dimensional systems strongly support
a power law $v^{-d}$,  with an exponent $d\in(0.327,0.375)$.
\end{abstract}

\maketitle

\noindent{\it Introduction}: Examples of quantum fluids such as liquid helium and 
ultracold quantum gases are characterised as either liquids or gases.  Remarkably,  given 
the right set  of circumstances,  ultracold quantum gases are able to change character
and condense into a liquid state of quantum droplets 
\cite{PhysRevLett.115.155302,PhysRevLett.117.100401}.
This new physical phenomenon opens up
an avenue into studying quantum liquids in a weakly interacting system
\cite{Semeghini_2018,Cheiney_2018,Cabrera_2018,PhysRevLett.122.090401}.  Furthermore,
it has been suggested \cite{Moss:2024mkc} that droplets are non-relativistic analogues of 
relativistic ``oscillon'' phenomena of interest in
elementary particle physics \cite{Makhankov:1979fc,Gleiser:1991rf,Gleiser:1993pt}.

The existence of liquid droplets requires a balance
between the attractive forces in an unstable atomic gas,  and some kind of stabilising
force.  One possible stabilising force arises from quantum vacuum fluctuations
in an atomic mixture
 \cite{PhysRevLett.115.155302,PhysRevLett.117.100401}.
The quantum forces are small,  but can play an important role when other mean
field forces almost cancel out.  Examples of droplet formation have been seen in
Bose-Einstein condensates with mixtures of two states with  intercomponent attraction and 
intracomponent repulsion,  where the vacuum energy contribution is repulsive
and stabilising
\cite{Semeghini_2018,Cheiney_2018,Cabrera_2018,PhysRevLett.122.090401}.

This paper aims to answer a simple question about the dynamics of multiple
droplet formation, what governs the size and separation of the droplets? 
It will be argued that,
changing the parameters of a suitable droplet-forming system through the
droplet forming regime at a rate $v$ leads to droplets of characteristic size 
$r_d\propto v^{-d}$.  This result is valid up to a limiting rate,  and depends
on there being a large number of droplets.  It is conjectured that the scaling
law is independent of the 
detailed form of the quantum corrections,  and consequently independent of dimension.  A 
semi-quantitative argument for the scaling law is presented,  and this is backed up 
by numerical simulations in two dimensions. So far, experiments have only been 
done with one or two droplets, where droplet size is determined by the
total number of atoms, and so new experiments would be needed to
test the new theoretical prediction.

The scaling law for droplet sizes resembles the Kibble-Zurek scaling law for structure
formation in second order phase transitions 
\cite{TWBKibble1976,Zurek:1985qw,delCampo2014},
which involves a power of the rate parameter. 
The physical origin of the scaling laws for the droplets has some similarities, 
in that there is a limitation on the size of structures that can form on some
freeze-out timescale.   In the thermal case,  this is set by the relaxation time of the medium.  In the case of droplet formation, 
which can occur in the absence of dissipation, the process is similar to the
``underdamped'' Kibble-Zurek effect \cite{Suzuki:2024upo}. However, an important difference
is that droplet formation can happen at zero temperature from purely quantum fluctuations
that cannot be modelled by a Langevin equation approach. There is also good evidence that a 
Kibble-Zurek type of scaling law
applies to other situations, for example some first order phase transitions in cold atom systems
\cite{Panagopoulos:2015uia,Kang_2017,PhysRevE.95.032124,PhysRevE.95.022127,2020Sci7292Q}.

\noindent{\it Droplet formation}: Consider a condensate of two atomic states
with atom number densities $n_\uparrow$ and $n_\downarrow$,  and scattering lengths 
$a_{\uparrow\uparrow}$,  $a_{\downarrow\downarrow}$ and 
$a_{\uparrow\downarrow}$.  The mixture becomes unstable
when $a_{\uparrow\uparrow}a_{\downarrow\downarrow}-a_{\uparrow\downarrow}^2<0$.  
The case of ``balanced" number  densities,
where $n_\uparrow/n_\downarrow=(a_{\downarrow\downarrow}/a_{\uparrow\uparrow})^{1/2}$,  
is favoured on energy grounds. The theory in this case can 
be described by a single mean field $\psi$ with total atom density $n$ and 
quantum-corrected energy density $V_{LHY}(n)$ \cite{PhysRev.106.1135}.
The field $\psi$ satisfies the Gross-Pitaevsky equation 
\cite{PhysRevLett.115.155302,PhysRevLett.117.100401},
\begin{equation}
i\hbar\dot\psi=-\frac{\hbar^2}{2m}\nabla^2\psi-\mu\psi+V_{LHY}'(n)\psi.\label{gpe}
\end{equation}
Stability of a homogeneous mixture with density $\bar n$ requires $V_{LHY}''(\bar n)>0$,
and the onset of instability is marked by a density parameter $n_d$ such that $V_{LHY}''(n_d)=0$
\footnote{The droplet density depends on atomic scattering lengths.
Instability can be approached either by increasing $n_d$ or decreasing the mean density $\bar n$.}.
The final density inside a droplet
is equal to $n_d$ up to a factor of Euler's constant.

The chemical potential parameter is a simple convenience that allows us
to fix  $\dot\psi=0$  for the homogeneous system, setting $\mu=V_{LHY}'(\bar n)$.
The stability of the homogeneous state is analysed in perturbation theory by introducing the
Bogoliubov modes, which have dispersion relation
\begin{align}
\omega^2=\frac{k^2}{2m}\left(\frac{\hbar^2k^2}{2m}+2\bar nV_{LHY}''\right).\label{disp}
\end{align}
Thus, unstable modes exist when $n_d>\bar n$ and $V''_{LHY}$ becomes negative,
marking the onset of droplet formation.  If the parameters change at a rate $v$
near the droplet transition,
then we can replace $V''_{LHY}$ by $-gvt$,  where $g$ is related to the third derivative
of $V_{LHY}$ and the instability sets in at $t=0$.  

Following Sazuki and Zurek \cite{Suzuki:2024upo},  finding the characteristic length scales 
can be broken down into two steps.  The first step is to find the freeze-out time when the instability becomes 
non-linear,  and the second step is to convert this into a length scale.  
Taking the dispersion relation (\ref{disp}) as a starting point, the Bogoliubov modes will satisfy a non-linear
equation of the form
\begin{equation}
m\delta\ddot\psi-gnk^2vt\delta\psi+O(\delta\psi^2)=0,
\end{equation}
in the small $k$ limit.
The $v$ dependence can be removed by introducing a time coordinate $\tilde t=v^{1/3}t$,
and rescaling $\delta\psi$ to remove any $v$ dependence in the $\delta\psi^2$ terms.
The solutions for any given $k$ mode are independent of $v$ and become non-linear 
at some fixed value for the rescaled time $\tilde t(k)$.

Returning to the dispersion relation (\ref{disp}),  there is a value of $k=k_{\rm max}$ for which 
the growth rate $\Im(\omega)$ reaches a maximum\footnote{The freeze-out time was found using a 
long wavelength approximation which is relaxed when
finding the length scale. This is consistent with the approach used in Ref. \cite{Suzuki:2024upo}},  
$k_{\rm max}\propto|V_{LHY}''|^{1/2}\propto(vt)^{1/2}$.
We may interpret the corresponding length scale as the droplet separation $r_s$,
$r_s\propto(vt)^{-1/2}$. Since the freeze-out time $t=v^{-1/3}\tilde t$, it follows that
\begin{equation}
r_s\propto v^{-1/3}.
\end{equation}
Knowing the mean density of the system $\bar n$, the density of a typical 
droplet $n_d$ and the separation between droplets
fixes the droplet radius $r_d$, $r_d=r_{\rm sep}\sqrt{\bar n/\pi n_d}\propto v^{-1/3}$.

The scaling rule governs droplet radii once they form. After formation, droplet mergers can have a significant
effect on their sizes, particularly  when their separation is small.  Furthermore, three particle collision losses 
can cause droplets to evaporate and reduce in size \cite{PhysRevA.69.023602}. 
Droplet mergers are included in the numerical results discussed below. 

\noindent{\it Simulation}: A two-dimensional numerical simulation has been used to test the theoretical
prediction for the droplet radii.  Near the unstable parameter range,  the Gross-Pitaevski equation (GPE)
becomes \cite{PhysRevLett.117.100401}
\begin{equation}
i\hbar\dot\psi=-\frac{\hbar^2}{2m}\nabla^2\psi-\mu\psi+g'n\psi+gn\left(\ln\frac{n}{n_0}-1\right) \psi.
\label{2dgpe}
\end{equation}
where $n_0$ is a fixed parameter.  Classical and quantum terms have been separated out
and given coefficients $g'$ and $g$ respectively. The formation of quantum droplets happens 
when an attractive classical force term ($g'<0$) balances a repulsive quantum term.
In practice,  such a two dimensional system can be embedded
into three dimensions with transverse dimension $a_\perp$,  in which case the parameters
$g'$ and $g$ are related to the scattering lengths by
\begin{equation}
g'=\frac{\sqrt{8\pi}\hbar^2}{ma_{\perp}}
\frac{a_{\uparrow\uparrow}a_{\downarrow\downarrow}-a_{\uparrow\downarrow}^2}
{a_{\uparrow\uparrow}+a_{\downarrow\downarrow}-2a_{\uparrow\downarrow}},\qquad
g=\frac{2\hbar^2}{m}\frac{|a_{\uparrow\downarrow}|^2}{a_{\perp}^2},\label{gprime}
\end{equation}
and $n_0$ is of order $1/(4\pi a_\perp a_{\uparrow\downarrow})$ 
\cite{PhysRevA.98.051604,pelayo2025phasesdynamicsquantumdroplets}.
Note that, after combining the potential terms as in Eq.  (\ref{gpe}),  
the droplet density parameter $n_d=n_0e^{-g'/g}$ inherits a very strong dependence on
the value of $a_{\uparrow\uparrow}a_{\downarrow\downarrow}-a_{\uparrow\downarrow}^2$.

If we define the healing length $\xi_0=\hbar/\sqrt{mgn_0}$ as a natural length scale,
and $\omega_0=g n_0/\hbar$,  then the quantitative features of the GPE are
governed by a single parameter $g'/g$. On the other hand, the initial conditions
depend on $\bar n/n_0$ and the fluctuations in the field
depend on a parameter $n_0\xi_0^2$. 

The GPE was solved using a Truncated Wigner approach,  where Gaussian random initial
conditions represent a vacuum or thermal state.  Initial conditions
were imposed in the regime $n_d<\bar n$.  The value of $g'$ was then ramped down slowly to put the system
into the range $n_d>\bar n$ when droplets are able to form. The rate $v$ is defined by
\begin{equation}
v=\left|\frac{d\ln n_d}{dt}\right|_{n_d=\bar n}=\frac{1}{g}\frac{dg'}{dt}.
\end{equation}
Each run was stopped at the same final atom density $n_d$. The droplet sizes
and separations were then extracted by implementing a search algorithm to
identify the individual droplets.

\begin{center}
\begin{figure}[htb]
\begin{center}
\scalebox{0.3}{\includegraphics{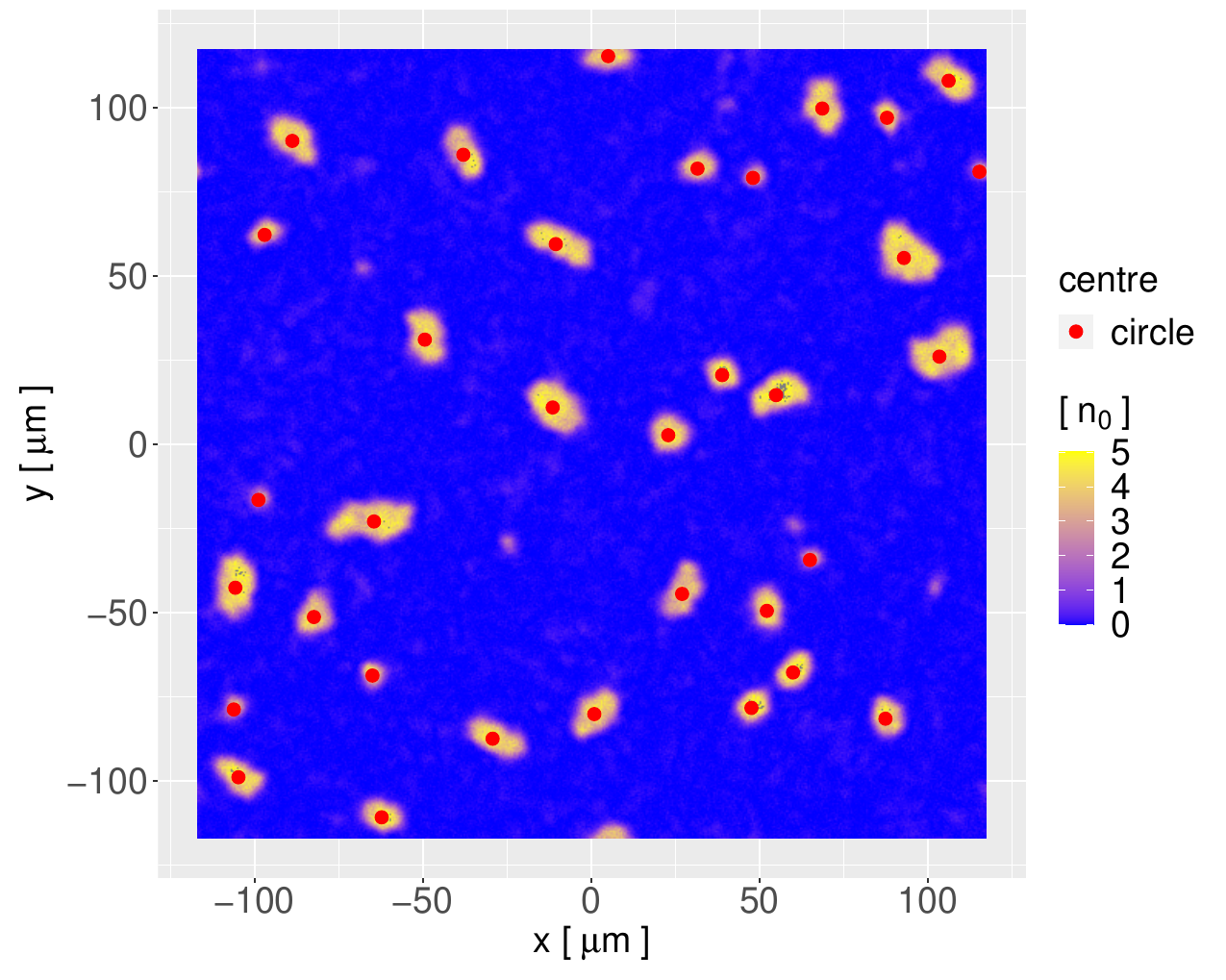}}
\scalebox{0.3}{\includegraphics{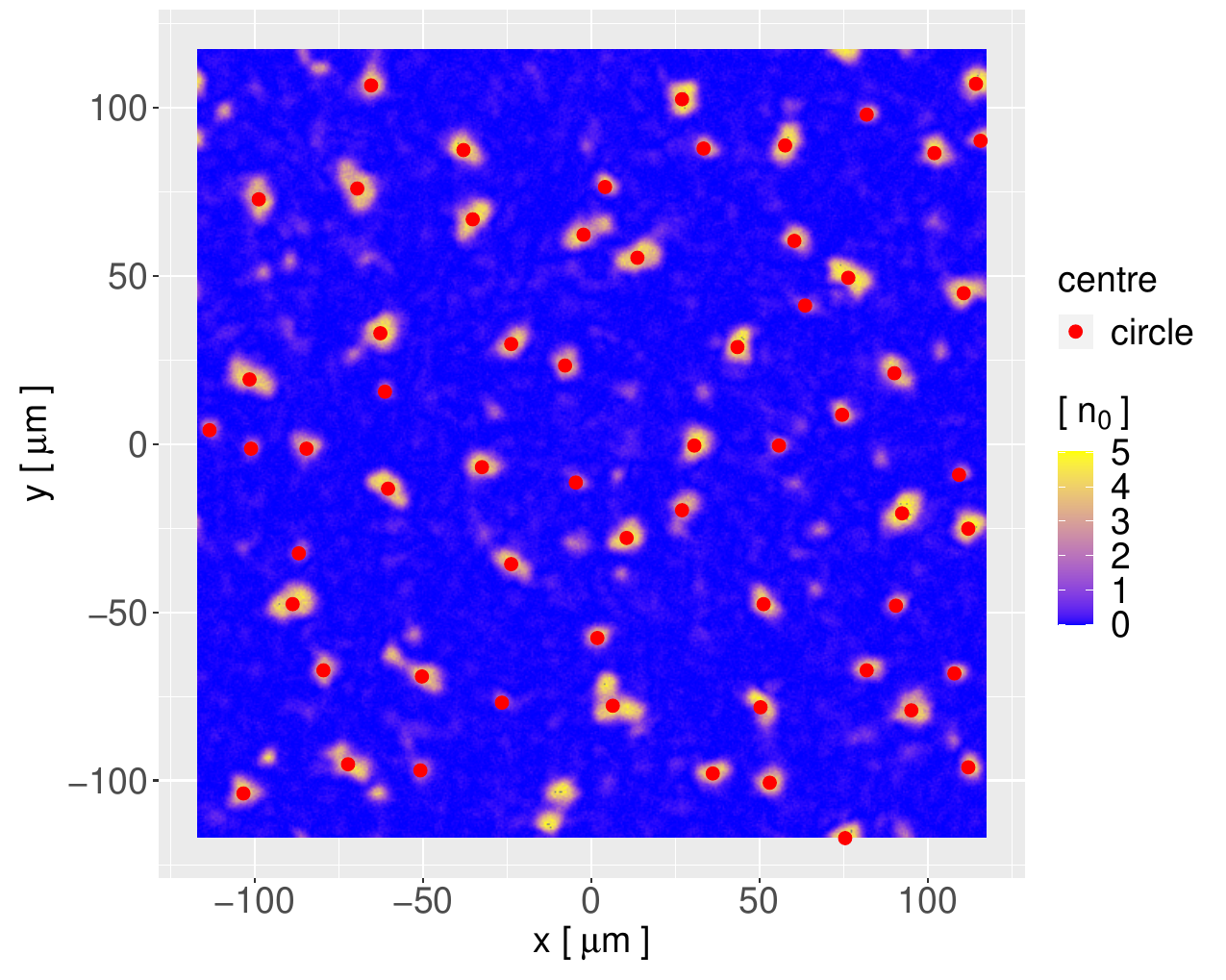}}
\end{center}
\caption{
The pictures show the density in a numerical (vacuum) simulation 
for the parameters given later in table \ref{tab:K}. The top picture
uses a lower rate  $v=0.004{\rm ms}^{-1}$, and the bottom
a rate four times larger $v=0.016{\rm ms}^{-1}$. The droplet  size is noticeably 
larger at the lower rate,  though mergers are present in both cases. 
Red dots show the droplet positions
found using a search algorithm.
}
\label{fig:sample}
\end{figure}
\end{center}

Two sample runs are shown in figure \ref{fig:sample} with two different values for the
rate $v$.  The figures clearly show that the droplets in the upper picture with a smaller
rate have a larger size. There is also a clear indication that droplets sometimes merge
into larger droplets,  and take on non-spherical shapes as a result. 

\begin{center}
\begin{figure}[htb]
\begin{center}
\scalebox{0.23}{\includegraphics{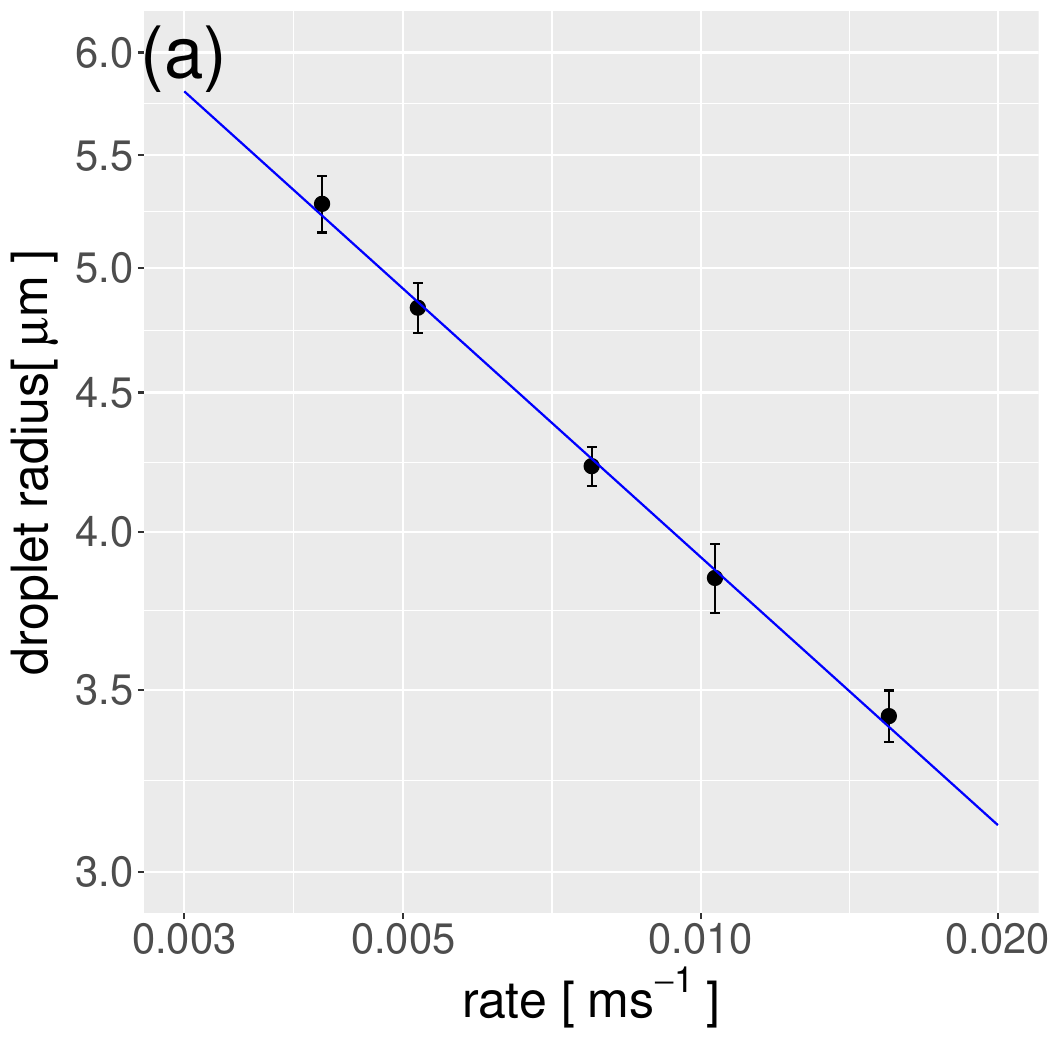}}
\scalebox{0.23}{\includegraphics{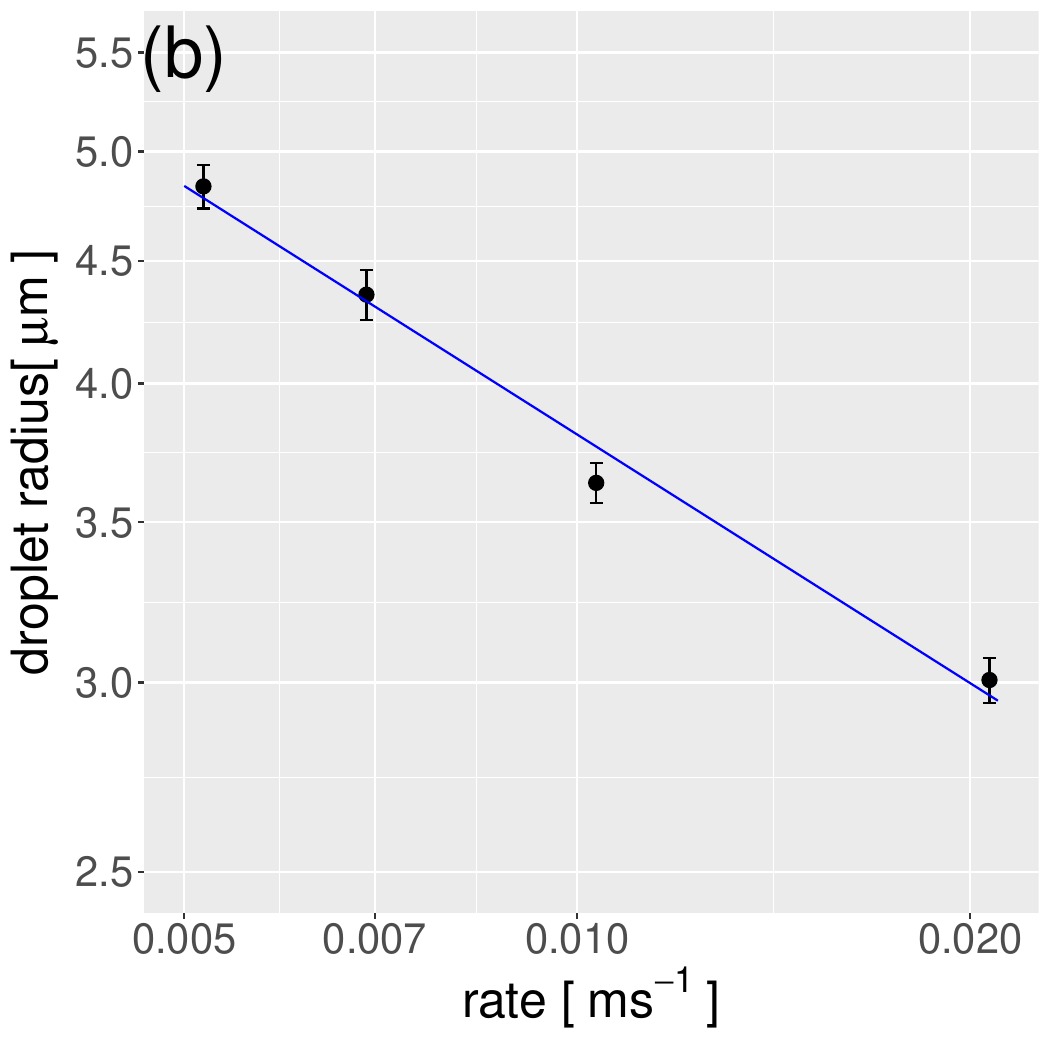}}\\
\scalebox{0.23}{\includegraphics{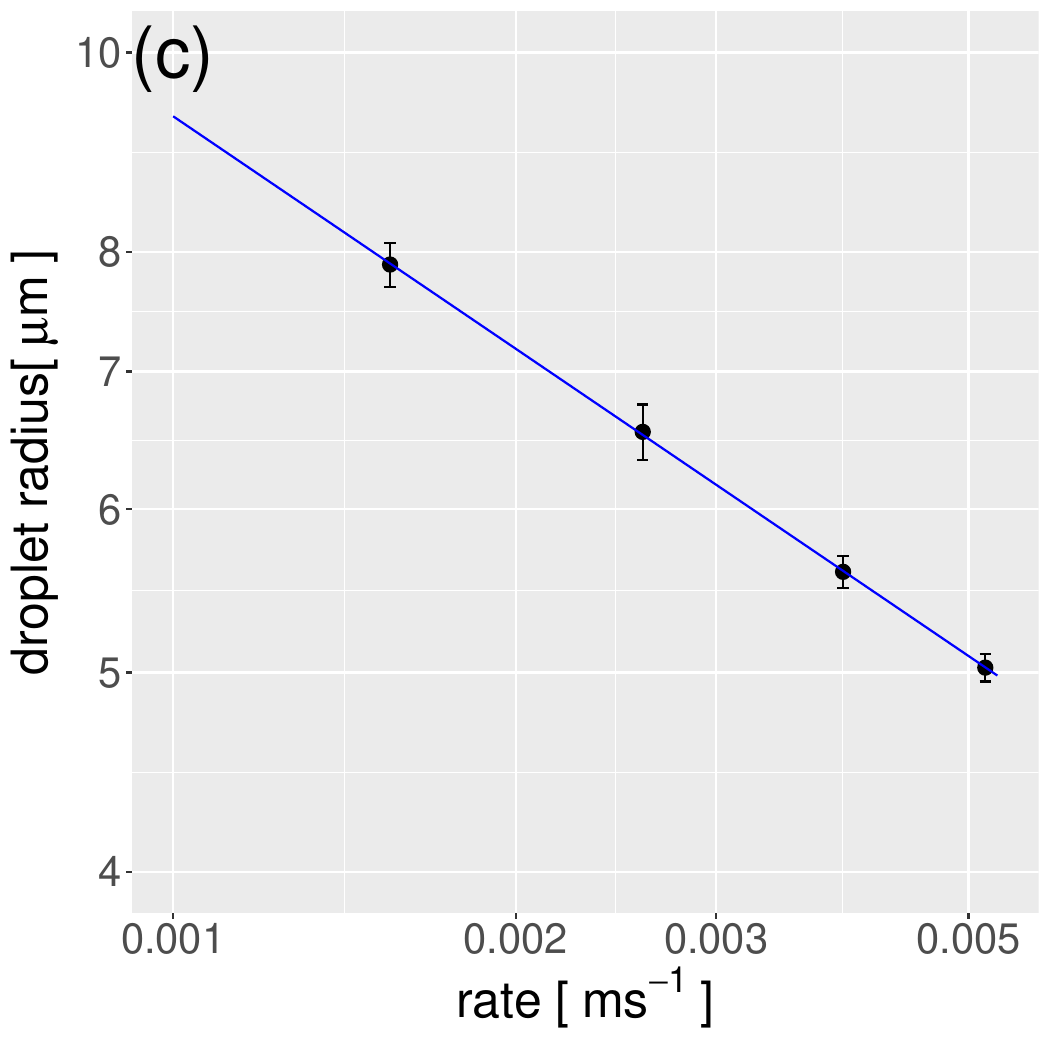}}
\scalebox{0.23}{\includegraphics{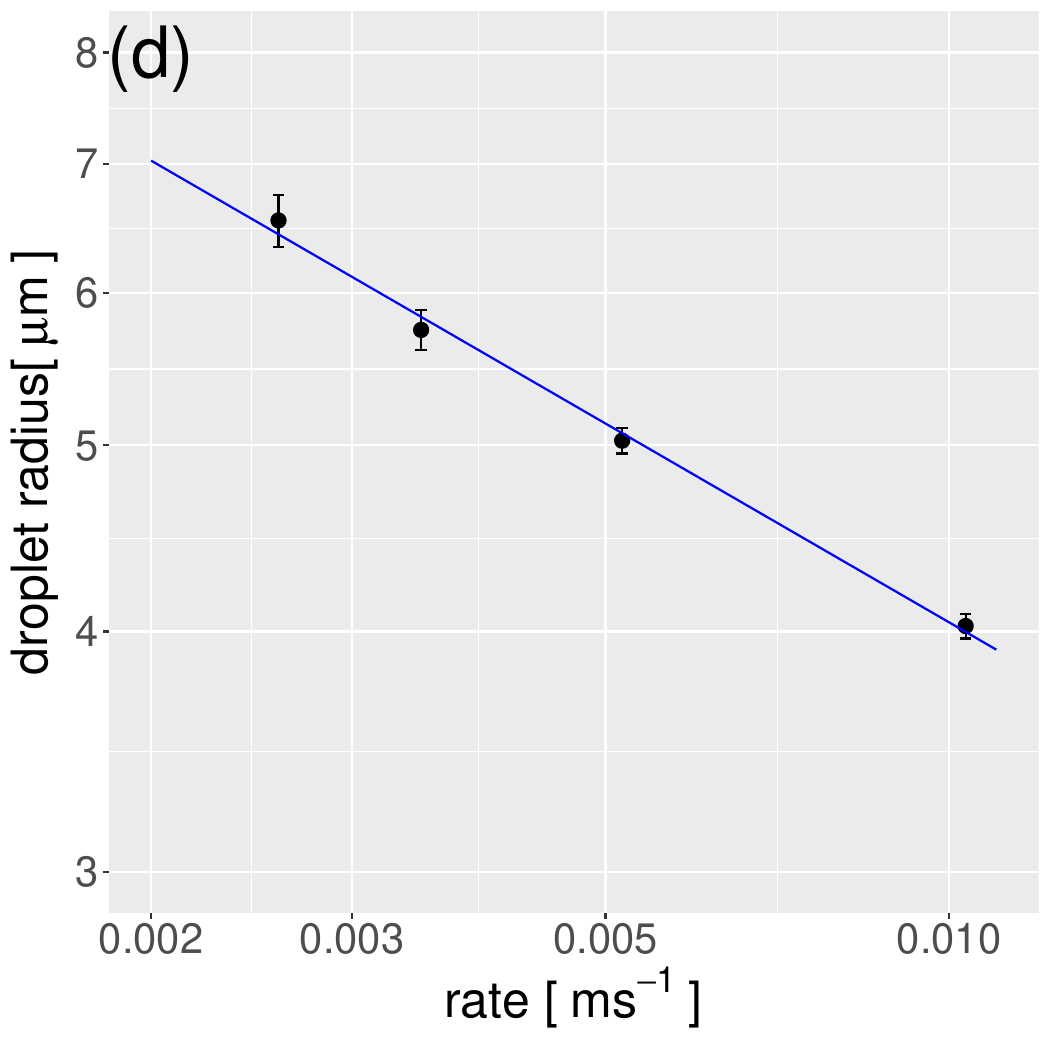}}
\end{center}
\caption{
Droplet sizes from the simulations plotted as a function of parameter evolution rate $v$.
Plots (a) and (b) use the parameters in table I, whilst plots (c) and (d) use half the
trapping frequency, $\omega_\perp=7.5\times2\pi {\rm kHz}$. In (a) and (c), the points
use different initial scattering lengths with fixed ramp duration. 
In (b) and (d), the initial scattering lengths are fixed whilst reducing the ramp duration. 
The lines show least squares fits $r_d\propto v^{-d}$,  with (a) $d=0.327$, (b) $d=0.345$, (c) $d=0.375$
and (d ) $d=0.343$.
The error bars show standard deviations over 10 runs.
}
\label{fig:sizes}
\end{figure}
\end{center}

The relation between droplet sizes and parameter evolution rates is shown in figure \ref{fig:sizes}.
The values of the parameters $g'/g$ and $n_0\xi_0^2$ differ by factors up to 4 between the plots to
show that the scaling relation goes beyond a trivial rescaling invariance of the GPE.
The data are excellent fits to power-law relations $r_d\propto v^{-d}$. The small variation in the
value of $d$ for large changes in regime is consistent with small effects due to droplet mergers.
However, the power $d\in(0.327,0.375)$ is consistent with the naive prediction $d=1/3$ given earlier.

\noindent{\it Experimental realisation}: The numerical simulations were based on a hypothetical
BEC mixture of potassium-39 in two hyperfine states $|1,0\rangle$ and $|1,-1\rangle$.  
At a field of around $56.85{\rm G}$ a Feshbach
resonance affecting the scattering length $a_{\uparrow\uparrow}$ satisfies the instability 
condition $g'=0$ \cite{DErrico2007,PhysRevLett.111.053202,Tanzi2018}.  
The parameters used for the simulations are given in table \ref{tab:K}.  
The initialisation was done with $B>56.85{\rm G}$ and $B$
was reduced to take the system through $g'=0$ at a rate $v$. 

An experimental realisation of the system would be subject to thermal effects.  In the first place,  the quantum potential
has been calculated using the quantum vacuum energy.  Thermal corrections would change the result 
if the thermal energy per particle exceeded the ground state energy of the Bogoliubov modes,  which happens at a temperature
$T_{LHY}$ given in table \ref{tab:K}.  A more exacting limit comes from the initial quantum fluctuations 
used in the Truncated Wigner approach.  Thermal fluctuations become important at temperatures above $T_{GPE}$. 

\begin{table}
\caption{\label{tab:K39constants}Physical parameters used in the potassium-39 simulations 
(for figures 2(a) and 2(b)).}
\begin{ruledtabular}\label{tab:K}.
\begin{tabular}{lll}
Parameter && Value \\
\hline  \\ [-3ex]
initial density&&$\bar n=50\mu m^{-2}$\\
system size&&$232\mu{\rm m}$\\
ramp time&&$543{\rm ms}$\\
rate&&$v\in(0.004,0.02){\rm ms}^{-1}$\\
critical field strength&&$B=56.85{\rm G}$\\
field time derivative&&$dB/dt\in(0.3,0.12){\rm mG}\,{\rm ms}^{-1}$\\
trap frequency&&$\omega_\perp=15\times2\pi{\rm kHz}$\\
droplet density scale&&$n_0=201\mu m^{-2}$\\
final droplet density&&$n_d=820\mu m^{-2}$\\
healing length&&$\xi_0=(\hbar/mg n_0)^{1/2}=2.34\mu m$\\
frequency scale&&$\omega_0=gn_0/\hbar=295{\rm Hz}$\\
LHY temperature limit&&$T_{LHY}=|g_{12}|n_d/k_B=279{\rm nK}$\\
vacuum temperature limit&&$T_{GPE}=\hbar\omega_0/k_B=2.4{\rm nK}$
\end{tabular}
\end{ruledtabular}
\end{table}

\noindent{\it Conclusions}: Ultra cold quantum gases condense into liquid quantum droplets under
certain parameter regimes.  In a many droplet system, the size distribution of these droplets is 
determined dynamically in the droplet transition.  It has been argued here that there is a refered 
droplet size and separation fixed in terms of the rate of change of parameters $v$ by 
$r_d\propto v^{-d}$,  where $d=1/3$. This is independent of the detailed nature of the system.

Numerical simulations show strong evidence for a power law, with $d\in(0.327,0.375)$.  
It would be interesting to pursue this further beyond two dimensions,  to examine the effects 
of thermal fluctuations, and to consider ``unbalanced" mixtures \cite{flynn2023}.  The proposed 
range of experimental parameters 
look challenging,  but further numerical
analysis may allow some movement in these values

The droplet transtion has similarities to other physical processes. In particular,  the droplets
are non-relativistic analogues of the ``oscillons'' in elementary particle physics
\cite{Makhankov:1979fc,Gleiser:1991rf,Gleiser:1993pt}.  Oscillons
play a role in phase transitions,  and are a candidate for dark matter \cite{Moss:2024mkc}. 
The study of quantum droplets
may shed light on these related areas of physics.

Data supporting this publication are openly available under a Creative Commons
CC-BY-4.0 Licence in \cite{DataArchive}

\acknowledgments
The author is grateful for helpful comments from Tom Billam, Matt Johnson, Markus Oberthaler, 
Yansheng Zhang and Mark Hindmarsh. This work is supported by the UK Science and Technology Facilities Council
[grants ST/T00584X/1 and ST/W006162/1].  The author is also grateful for hospitality of the Perimeter 
Institute and the University of Helsinki where part of this work was carried out.  
Research at Perimeter Institute is supported in part 
by the Government of Canada through the Department of Innovation, Science and Economic 
Development and by the Province of Ontario through the Ministry of Colleges and Universities.

\bibliography{paper.bib}

\section*{Supplementary Material}

\noindent{\em Renormalisation} It is instructive to analyse the renormalisation of the quantum vacuum energy
in two dimensions which has been used in the main text. We start with separate number densities in the two
BEC components $n_\uparrow$ and $n_\downarrow$. The quantum vacuum energy density $\Delta V$ 
depends on the two-dimensional couplings $g_{ij}$.
It requires a momentum cutoff $\kappa$ to obtain a finite result\cite{PhysRevLett.117.100401},
\begin{equation}
\Delta V=\frac{\hbar^2}{m}\kappa^4
+\frac{m}{8\pi}\kappa^2\sum_\pm c_{\pm}^2+
\frac{1}{8\pi}\frac{m^3}{\hbar^2}\sum_\pm c_{\pm}^4\ln\frac{m^2c_\pm^2}{\hbar^2\kappa^2}
\label{dVa}
\end{equation}
where
\begin{equation}
2mc_\pm^2=g_{\uparrow\uparrow}n_\uparrow+g_{\downarrow\downarrow}n_\downarrow\pm
[(g_{\uparrow\uparrow}n_\uparrow-g_{\downarrow\downarrow}n_\downarrow)^2
+4g_{\uparrow\downarrow}^2n_\uparrow n_\downarrow]^{1/2}
\end{equation}
The first term in $\Delta V$ is the usual energy density divergence, and the second can be absorbed into
a homogeneous phase rotation. The cutoff dependence $\ln\kappa^2$ in the final term depends on the combination
\begin{equation}
\sum_\pm c_{\pm}^4=\frac{1}{2m^2}\left(g_{\uparrow\uparrow}^2n_\uparrow^2+
g_{\downarrow\downarrow}^2n_\downarrow^2+
2g_{\uparrow\downarrow}^2n_\uparrow n_\downarrow\right).
\end{equation}
This has the same form as the classical scattering terms, and can be absorbed by renormalisation
of the couplings. We put a cutoff dependence into the ``bare" couplings and define the renormalised 
couplings $g_{ij}^R$ by
\begin{equation}
g_{ij}(\kappa)=g_{ij}^R+\frac{1}{8\pi}\frac{m}{\hbar^2}(g_{ij}^R)^2\ln(\kappa^2/\mu_R^2).
\end{equation}
The renormalisation scale $\mu_R$ must be introduced for consistency, and it replaces 
the cutoff $\kappa$ in Eq. (\ref{dVa}). In some respects, $\mu_R$ is an arbitrary scale. However, other 
quantum corrections will depend on the renormalisation
point, and some choices are better than others at suppressing these. In this example, one obvious source of
corrections is the third dimension, and we can use results that include a third, compact, dimension
to guide the choice of renormalisation point 
\cite{PhysRevA.98.051604,pelayo2025phasesdynamicsquantumdroplets}. 
These results contain a term similar to (\ref{dVa}), with $\mu_R\sim 1/a_\perp$, which 
suggests choosing this as the renormalisation point.

Returning to the balanced case for the number densities, then both 
$c_-^2$ and the classical potential are 
$\propto g_{\uparrow\uparrow}g_{\downarrow}-g_{\uparrow\downarrow}^2$,
which is small close to the instability point.
The quantum term is also reduced by extra factors of $1/a_\perp$, so we neglect $c_-^4$, but keep
the classical terms and $c_+^4$. By choosing $\mu_R$ as indicated above, we get the simple form
for the GPE given in the main text.

\noindent{\em Details of the numerical simulation} The numerical scheme integrates the projected GPE (PGPE)
with a variable timesstep integration scheme. The projection is introduced for de-aliasing, removing 
modes with $k$ values larger than half the lattice cutoff. The runs used to produce the resuts
have grid size $512\times512$, but the results are unchanged when the grid is increased
to $1024\times1024$. 

Following the truncated Wigner approach, multiple runs start from initial conditions drawn from a gaussian 
distribution which replicates the symmetrised two-point correlation function of the quantised field
fluctuations. 
The runs start in a regime with $g'\ne 0$, and the correlation function
\begin{equation}
\langle \{\delta\bar\psi(k)\delta\psi(k')\}\rangle=\frac12\frac{K+g'\bar n}{\sqrt{K(K+2g'\bar n)}}\delta_{kk'},
\end{equation}
where $K=\hbar^2 k^2/2m$ and $\{\}$ denotes the anticommutator. After rescaling, 
the correlations depend
on $\bar n/n_0$, with an overall factor  $1/n_0\xi_0^2$. Numerical simulations were
also run with white noise initial conditions $\langle \bar\psi(k)\psi(k')\rangle=\frac12\delta_{kk'}$
with no significant change to the results.

\begin{center}
\begin{figure}[ht]
\begin{center}
\scalebox{0.25}{\includegraphics{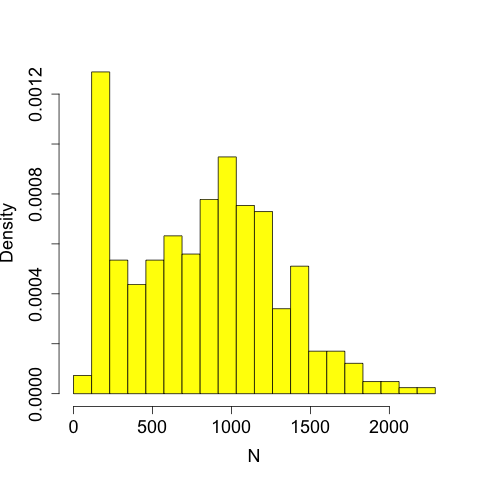}}
\end{center}
\caption{
The histogram shows the probability distribution of atom numbers in each lump.
(Compiled from five separate runs with $v=0.01{\rm ms}^{-1}$.)
}
\label{fig:drops}
\end{figure}
\end{center}

The value of $g'$ in each run varies linearly for the ramp duration and ends at a fixed value of $g'/g$,
so that the final droplet density $n_d$ is fixed, The droplets are identified using a search algorithm.
Figure \ref{fig:drops} shows the atom number distribution over the droplets for a sequence of runs with the
same parameters.  There is a distinct  separation into liquid droplets, on the right, with constant internal 
density, and small solitonic style lumps with radii of order the healing length.
These small lumps are removed before testing the scaling law by putting a threshold $n_d/2$ on the
internal density. The mean radius $r_d$ is defined from the atom numbers and density, and the
separation $r_{\rm sep}$ from the number of droplets per unit area. Both satisfy identical scaling laws.

\noindent{\em Details of the experimental proposal} 
 The scattering length $a_{\uparrow\uparrow}$ of the hyperfine state 
 $|1,0\rangle$ of potassium-39 varies strongly 
with magnetic field around $57{\rm G}$, and 
$a_{\uparrow\uparrow}a_{\downarrow\downarrow}-a_{\uparrow\downarrow}^2$ changes sign
from positive (stable) a little above $56.85{\rm G}$ to negative below.
The rate $v$ used in the text can be related to the rate of change of the magnetic field
using Eq. (\ref{gprime}) and scattering lengths from the literature 
\cite{DErrico2007,PhysRevLett.111.053202,Tanzi2018},
\begin{equation}
v=\frac{1}{g}\frac{dg'}{dt}\approx
131\frac{a_{\perp}}{1\mu m}\frac{dB}{dt}.
\end{equation}
Broadening the range of some of the parameters in table \ref{tab:K} is restricted
by a number of constraints.
Primary requirements are  $r_{\rm sep}>r_d$, so that
 the droplets are well-separated. A second requirement $r_d>\xi_0$,
so that the radius is larger than the thickness of the wall of the droplet. 
A third requirement is that the separation should be less than the overall size
of the system. A last inequality  $\xi_0>a_\perp$, is required for an 
effective two-dimensional theory.  This last constraint may
be relaxed when investigating systems which are not two dimensional.

\vfill

\end{document}